\documentclass{iopart}

\usepackage{latexsym,graphicx} 
\begin{document}

\title[The effects of material and surface geometry on vdW forces]{Non-additive behavior of van der Walls dispersion forces due to material and surface geometry effects }

\author{C E Rom\'{a}n-Vel\'{a}zquez and Bo E Sernelius}

\address{Dept. of Physics, Chemistry, and Biology, Link{\"o}ping
University, SE-581 83 Link{\"o}ping, Sweden}
\ead{roman@ifm.liu.se}
\begin{abstract}
We present a series of calculations of van der Waals (vdW) forces that show
non-additive behavior.  The results reveal effects of geometrical dependences
of the dispersion forces, that are in strong contradictions to the results
from additivity approaches like the Lennard-Jones (LJ) pairwise interaction
and the proximity force approximation (PFA).  For the simple geometries
treated here deviations appear when the effects of the materials are taken
into account.  We obtain these results with a general and simple numerical
method that allows for a detailed study of the vdW interactions.  The
method is also accurate and fast, and due to its generality makes the study
of interactions between three-dimensional (3D) objects, with arbitrary
geometries and configurations, feasible.  We present results for normal,
lateral and rotational forces.
\end{abstract}

%\maketitle
In recent years a perfection of chemical and physical methods has allowed
for building so called nano-systems, in which the distances and sizes of
the objects are of the order of tens of nanometers.  One of the most
interesting phenomena that emerge at this scale are the Casimir and vdW
forces known as dispersion forces.  They arise as a consequence of the
quantum fluctuations of the electromagnetic fields.  These forces have been
measured experimentally at different conditions, and different theoretical
methods have been developed for their study.  The basic understanding of
these interactions has increased in recent years but much of the physics of
these phenomena need to be better known. The experimental success in
measuring the Casimir forces \cite{Lamoreaux} has opened up for an animated
discussion and research about the possibility to develop nano-mechanical
devices \cite{nems}.  Now that subject has survived the
speculative phase and several simple mechanical devices have been developed
\cite{nemscapasso}.  In order to understand, predict and design these
devices it is necessary to study the dispersion forces
between objects with different geometrical shapes, of different materials,
and in different configurations.  In doing so it is common to use more
simple and accessible approximate methods that rely on additivity
approximations.  They are based on an assumption of a sort of additive
interaction; the force is calculated as the sum of contributions from
different parts of the system.  As a consequence, one can only expect an
increase in the interaction of the system when the dimensions of the objects
are increased.

Here we develop a series of calculations of vdW forces for simple 3D
objects. We treat cylinders, with circular and square cross sections, in
different arrangements, interacting with each other or with a substrate;
we study normal, lateral and rotational forces. We show that when the effects
    of surface geometry and material are taken into account even these
    simple geometries show a non-monotonic and non-additive behavior.
We have performed these calculations with the use of a
new numerical method that is fast, easy to implement, and
provides results with high precision. Its generality make it possible to
apply on 3D finite objects with arbitrary form and orientation.

One of the more common methods based on additivity approximations has been
formulated since the beginning of the study of the vdW interactions.  It
consists in the summation of the contributions from pairwise LJ
interactions between the atoms of the different objects.  This approach
has, for example, recently been used to calculate the vdW forces between
fullerenes and nanotubes \cite{Girifalco}.  For fullerenes the results have
been compared with more realistic calculations and deviations of around
10\% have been found \cite{Girard}.  Another additivity approach is the
PFA. One discretizes the surface of the interacting objects in small
parallel plane surfaces and treats the interaction between them like that
of two infinite slabs.  A sum of the forces between interacting pair planes
gives the force between the objects.  Rigorous theoretical models have
confirmed the predictions of this model in the vdW case for dielectric
spheres and cylinders \cite{Langbeinbook}.  In experiments agreements
within 1\% have been claimed \cite{precision}.  More recent
theoretical models have
tried to determine the applicability of the PFA
\cite{Periodic,PFAlimits} and external criterions for the formulation
of the PFA have been postulated to extend its applicability
\cite{Mazzitelli,PFAlimits}.

   From the initial formulation of the Casimir force \cite{Casimir} one has
come to realize the strong influence the geometry has on the properties of
the system.  Casimir showed that an interaction energy between two flat
infinite plates arises from the variation of the zero point energy
relative its value at infinite separation \cite{Casimir},
\[
\mathcal{E}(z)=\frac{\hbar}{2}\sum_{s}\omega_{s}(z)-\frac{\hbar}{2}\sum_{s}\omega_{s}(z\rightarrow\infty),\]
where $\hbar$ is Planck's constant and $z$ the distance between the
plates.  In the general problem there is a dependence on the surface
geometry and materials of the system due to the boundary
conditions that the electromagnetic fields need to satisfy.  In the
idealized Casimir
system only vacuum modes exist.  For real metal objects both vacuum modes
and surface modes contribute.  The vacuum modes dominate the force in the
retarded limit, the Casimir limit, and the surface modes in the
non-retarded limit, the vdW limit.

Several theoretical methods have dealt with the problem of calculating the
dispersion forces from the electromagnetic resonances of the system.  The
major part of them have been formulated for ideal perfect metals
\cite{revmilton}.  They have given an
important insight into the physics of the phenomenon.  However their
results are not directly applicable to real systems.  It has been shown
that effects of the actual materials are important.  In particular the
surface plasmons are essential in the description of the phenomenon in the
situation of close proximity of the objects \cite{SerneliusPRB}.  The
perfect metal model has no vdW range, which means that the results from this
model only apply to experiments on actual micro-devices if the objects are
far apart.  Only recently a multipolar scattering formalism has been 
formulated for
calculation of the interaction between dielectric objects
\cite{Casphe}.  These kind of methods give results with great accuracy over
a wide range of distances for highly symmetric objects like spheroids or
cylinders
\cite{mulsphe,Mazzitelli,cylplate}. More complex
geometries give rise to a more complex description with a slower
computation convergency and low accuracy.  A general numerical method for
the calculations of electric fields for arbitrary geometries and material
has been developed \cite{Capassocomp}.  However a deeper knowledge in
advanced numerical methods is required in order to obtain results with
reasonable accuracy and computational time: this hampers its general use.

A more simple but also more accurate theoretical method is required to go
beyond the more basic geometries and configurations that characterize the
recent results; it should also be able to deal with the effects of
different materials.  Here we are interested in and focus on systems with
distances and sizes of tens of nanometers.  From the practical point of
view, given the great number of nano-systems, existing and in progress,
this focus does not constitute a severe limitation on the use of the model.
However, much of the qualitative behavior found here is expected to be
found also in slightly bigger systems where the retardation effects only
start to be important.

In what follows we present a general theoretical method based on the
solution of a surface integral equation.  This method has been used in the
past to solve problems of scattering.  We have here adapted it to the
calculation of vdW forces.  The geometries of the objects we study here are
simple but thanks to this method it has been possible to develop novel
calculations of lateral and rotational forces, from which very few results
exist in the literature.

\section{Formalism}

Consider a finite object, immersed in the vacuum, of arbitrary form defined
by the surface $\Sigma$ and made of a non magnetic material with dielectric
function $\varepsilon\left(\omega\right)$.  In the non-retarded limit the
behavior of the electromagnetic fields is determined solely by the electric
potential.  The solution of the 3D Laplace equation on the boundary of the
object satisfies the Fredholm integral equation \cite{Phillips}
\[
\phi^{0}(\mathbf{r})=\frac{\varepsilon\left(\omega\right)+1}{2}\phi(\mathbf{r})+
\frac{\varepsilon\left(\omega\right)-1}{4\pi}\int_{\Sigma}dS\mathbf{n}'\cdot\nabla 
G\left(\mathbf{r},\mathbf{r}'\right)\phi(\mathbf{r}'),\label{eq:
intequa}\]
where $G=1/\left|\mathbf{r}-\mathbf{r}'\right|$ is the Green's
function of free space, $\mathbf{r}$ and $\mathbf{r}'$
points on the surface, $\Sigma$, of the object, and $\phi^{0}$ is the
electric potential derived from exterior sources.

The properties of the kernel of this integral equation have been studied
in the old potential theory \cite{Kellogs}. Due to its properties
it is possible to apply the theory of the Fredholm integral equation and
the different analytical and numerical methods to its solution. Here
we deal with the most simple numerical method, known as boundary element
method.  We have adapted it to obtain the resonances of the system,
i.e., to find solutions different from zero when the applied external
field, $\phi^{0}$, is identically zero.
The method consists in the discretization
of the surface $\Sigma$ in $N$ small planes of sizes $\Delta
s_{i}$ in the positions $\mathbf{r}_{i}$
with normal unit vector $\mathbf{n}_{i}$ and potential
$\phi_{i}$. A homogeneous system of linear equations is obtained: $
\sum_{j}R_{i,j}\left(\omega\right)\phi_{j}=0,$
with \[
R_{i,j}\left(\omega\right)=2\pi\frac{\varepsilon\left(\omega\right)+1}{\varepsilon\left(\omega\right)-1}
\delta_{i,j}+(1-\delta_{i,j})\frac{\mathbf{n}_{j}\cdot\left(\mathbf{r}_{i}-\mathbf{r}_{j}\right)}
{\left|\mathbf{r}_{i}-\mathbf{r}_{j}\right|}\Delta
s_{j}.\]
The zero point energy of the system is given by
\[
U\left(\omega\right)=-\frac{\hbar}{i4\pi}\int_{-i\infty}^{i\infty}d\omega\log
\det R_{i,j}\left(\omega\right) . \]
Note that there is no restriction on the shape and connectivity of
$\Sigma$.  Therefore, we can consider a surface that consists of two
arbitrary non overlapping surfaces $\Sigma_{1}$ and $\Sigma_{2}$ with
arbitrary positions and orientations.  It is possible to study an object
above a substrate by letting one of the objects be the mirror image of the
other.  The total energy of the system is obtained by numerically
calculating the integral of the complete system and subtracting the value
obtained considering the objects independently.  The computational time
goes like $\mathcal{O}\left(N^{2}\right)$, but $N$ grows slower than in
methods like that in \cite{Capassocomp}, since our method is based on a
discretization of a surface.  With this method it is possible to, within a
few hours, obtain results for the interaction energy with a precision of
1\% at distances down to 5\% of the diameter of the smallest of the
objects.  This conclusion we draw from comparisons with high precision
calculations based on methods in \cite{mulsphe}.

\section{Calculations and Results}

\begin{figure}
\center
\includegraphics[width=8cm]{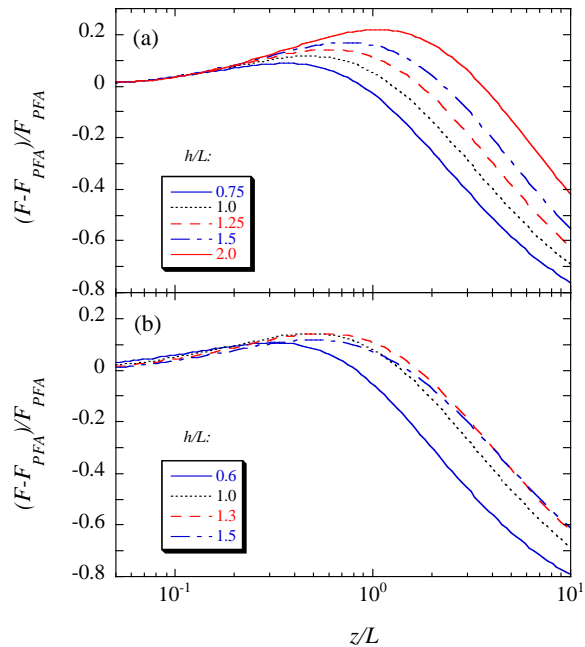}
\caption{\label{fig: I}Normal forces between a finite gold cylinder and a
gold substrate; (a) circular cross section with diameter $L$; (b) square cross
section with side length $L$.}
\end{figure}

In figure (\ref{fig: I}) we show results from calculations of the
interaction between a gold cylinder and a gold substrate.  The base of the
cylinder is kept parallel to the substrate and at the distance $z$.  Panel
(a) is for cylinders with circular cross section of diameter $L$ and panel
(b) for square cross sections with side length $L$.  The curves are for
different heights, $h$.  We have plotted the relative difference between
our results and those from PFA. The PFA result is the force per unit area
between two half spaces times the base area; it does not depend on the
height at all.

First we observe in both panels that all the curves tend to zero
in the small separation limit which means that all of them present
the behavior of two
parallel plates in the limit.  At large distances a volume dependence is
observed corresponding to dipolar interactions.  This means that a larger
cylinder has a stronger interaction with the substrate.  At intermediate
distances we observe that in the circular cylinder case, panel (a), the
bigger cylinders always show a stronger vdW force.  One can expect this
behavior in an interaction that depends on the quantity of material, as is
the case, for example, of an additive pairwise interaction.  However the
square cylinders in panel (b) show a completely different behavior: a
non-monotonic relationship between the forces and the volumes of the objects.
There are regions where an increase of the volume leads to a decrease
in the force.
For example at a
distance smaller than $0.8L$ the square cylinder with height $1.5L$ is
expected to show a greater force that the cylinder with $h=L$, but it
does not.  This indicates a sort of resonant behavior that depends on the
geometry.  These results agree with the geometry dependence of the resonant
modes of the electromagnetic field.

This lack of direct relation between the force and volume of the object is
also observed in the case of lateral forces.
\begin{figure}
\center
\includegraphics[width=8cm]{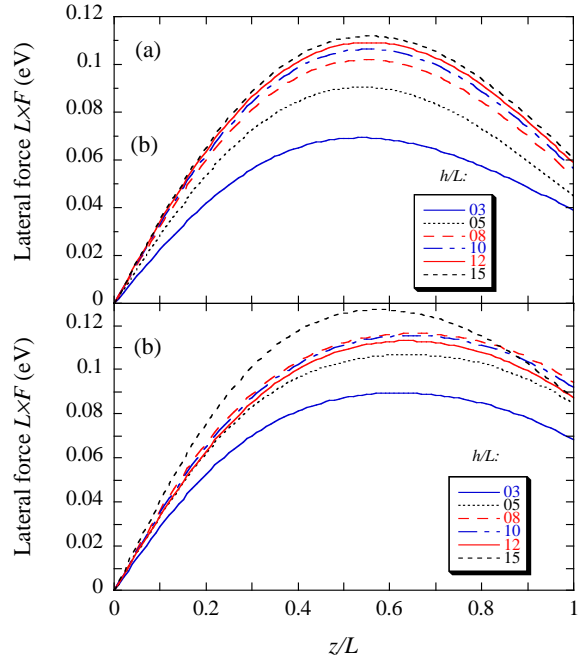}
\caption{\label{fig:lateral}Lateral forces between two gold cylinders kept
at the vertical separation $L/2$ and moved horizontally the distance $z$;
(a) circular cross section with diameter $L$; (b) square cross section with
side length $L$.}
\end{figure}
In figure (\ref{fig:lateral}) we show results from calculations of lateral
forces between pairs of finite cylinders.  Initially the cylinders are
considered aligned above each other with parallel bases at a distance
$d=L/2$ from each other.  Then there is a horizontal, lateral displacement
$z$ parallel to one of its sides of length $L$ (in the case of cylinders
with square cross section).  The energy is calculated at different
positions $z$ and the lateral force is obtained through numerical
derivation.

We observe that in the case of circular cylinders, panel (a) the objects
with greater height show a greater lateral force than the objects with
smaller height.  Again this is the kind of behavior that one can expect
through an additive interaction.  However this is not always the case when
one considers finite square cylinders.  In panel (b) we present results for
pairs of square cylinders of the same height.  The lateral force increases
in general with the height but for the cylinder of height $1.5L$ the force
is smaller than for cylinders of heights $1.0L$, and $1.2L$ in a region of
lateral displacements.  Note, the wide region of linear behavior in the
lateral force, which implies a harmonic oscillation around the completely
aligned equilibrium.  The PFA predicts a constant lateral force; it has
been shown that this happens in the perfect metal case for periodic 2D
steps for small normal separation \cite{Periodic}.  However the
non-constant linear form dominates the behavior of the dispersion forces
when the objects are in close to alignment with each other even if they are
in close proximity.  This is a geometric dependence of the interactions
between the borders of the objects that the PFA is not able to take into
account.

\begin{figure}
\center
\includegraphics[width=8cm]{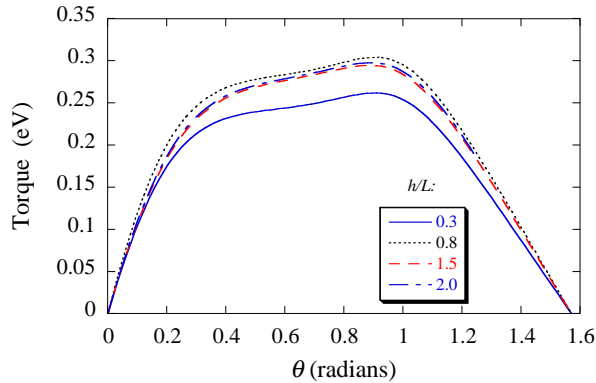}
\caption{\label{fig:rotational}Torque between two gold cylinders with
rectangular cross section, $L \times 2L$, and height $h$, kept at the
vertical separation $L/2$.}
\end{figure}

In figure (\ref{fig:rotational}) we present results for rotational forces,
or torques, between rectangular finite cylinders.  The base is a rectangle
with the longer side twice the length of the shorter.  Each curve
corresponds to a different height.  Initially the cylinders are considered
in their position of minimum energy where they are with their bases
parallel and completely aligned.  They are rotated around a perpendicular
axis through the center of the bases.  We calculated the energy at
different angles and found the rotational force by numerical
differentiation We observe that the rotational force is not stronger for
the greatest cylinder.  The cylinder with height $0.8L$ is of a sort of
resonant height that gives the maximum rotational force.  Also here, like
in the lateral force, it is easy to see that the PFA will predict a
constant rotational force in the proximity of the close alignment.
However, one expects the non-constant linear behavior of the force close to
the equilibrium position, independently of the proximity of the objects.

A non-monotonic behavior of the Casimir forces was also found in a recent
work on perfect metal systems \cite{Capassocomp}.  It involves complex
interactions between walls and objects.  Here we have shown that
non-monotonic and non-additive behaviors can be seen in more simple systems
when the effects of the materials are taken into account.  These results
are in conflict with those expected in PFA and from pairwise LJ
interactions.  Here, we have furthermore presented a method that
constitutes a useful and versatile tool for the calculations of vdW forces
in different configurations; it is able to deal with objects of different
geometries and in different configurations; it gives fast and high
precision calculations.  The precision can be improved with an optimal
discretization of the surface and by exploiting the symmetries of the
systems, like axial and translational symmetries.  The method can be
generalized to deal with systems that involve different materials.

\ack
This research was sponsored by EU within the EC-contract No:012142-NANOCASE
and support from the VR Linn\'{e} Centre LiLi-NFM and from CTS is
gratefully acknowledged.

\end{document}